\documentclass[pre,twocolumn,showpacs,floatfix]{revtex4-1}
\usepackage{psfrag}
\usepackage{amssymb,amsmath,amsthm}
\usepackage[dvips]{graphicx}
\usepackage{verbatim}
\usepackage[colorlinks=true,linkcolor=blue,citecolor=blue]{hyperref}

\begin{document}

\title{
Synchronization
of globally coupled two-state stochastic oscillators with a
state-dependent refractory period}

\author{Daniel Escaff$^{1}$, Upendra Harbola$^{2,3}$
and Katja Lindenberg$^{2}$} 
\affiliation{ $^{1}$Complex Systems Group, Facultad de Ingenier\'\i a y
Ciencias Aplicadas, Universidad de los Andes, Av. San Carlos de
Apoquindo 2200, Santiago, Chile\\
$^{2}$Department of Chemistry and Biochemistry and BioCircuits Institute,
University of California San Diego, 9500 Gilman Drive, La Jolla, CA 92093-0340\\
$^{3}$Department of Inorganic and Physical Chemistry, Indian Institute
of Science, Bangalore, 560012, India}

\begin{abstract}

We present a model of identical coupled two-state stochastic units each of which in
isolation is governed by a fixed refractory period. 
The nonlinear coupling between units
directly affects the refractory period,
which now depends on the global state of the system and can therefore itself become
time dependent.  At weak coupling the array settles into a quiescent stationary state. 
Increasing coupling strength leads to a saddle node bifurcation, beyond which the quiescent state 
coexists with a stable limit cycle of nonlinear coherent oscillations.  
We explicitly determine the critical coupling constant for this transition.

\end{abstract}


\maketitle

\section{Introduction}

Many dynamical processes in nature can in the first instance be modeled as
on-off processes.  Examples include firing of neurons~\cite{ElsonPRL98,Murray},
up-down spin systems, and blinking phenomena~\cite{fran}. Although the dynamics of such 
on-off single systems is simple, when coupled together in large numbers
they are known to exhibit complex global dynamical behaviors
such as synchronization, in which the entire system evolves as a single unit. 
The study of the synchronized dynamics of two-state networks is essential not only
as a problem of fundamental scientific importance~\cite{Zhou}, but also as a tool to 
help us understand many physiological processes~\cite{GlassNature2001}. For
example, our cardiac system is a giant network of synchronized
oscillators.  A vast number of systems
in physics and engineering also share such synchronization
features~\cite{Pikovsky,StrogatzPRL96}.  

Coupled two-state units offer the simplest example of coupled units 
of discrete states.  Such discrete models are
useful for the study of phase synchronization of
stochastic coupled oscillators~\cite{Wood,Wood2} and of excitable media~\cite{Prager,Kouvaris}.
They can be invoked in ecological applications~\cite{BerandIJBC2000} where populations
of various species interlinked through the
food chain exhibit spatial and temporal synchronization.
For sufficiently weak coupling between oscillators,
the system as a whole typically reaches a stationary state where on average the dynamics
ceases to exist. However,
for relatively strong coupling each unit is driven away from its equilibrium,
and the system as a whole may undergo a transition to a non-stationary state.

Phase synchronization on networks has most commonly been analyzed in terms of a Hopf 
bifurcation \cite{Wood,Wood2,Prager,Kouvaris}, that is, as a linear instability
of the stationary state. Synchronization may also occur via other bifurcation mechanisms
such as, for
instance, saddle node or homoclinic bifurcations~\cite{StrogatzLib}. 
A sub-critical Hopf bifurcation may, for example, involve a
prior saddle node which can not be captured within the linear analysis. 
Such sub-critical scenarios have been observed in a variety of different
contexts including the Kuramoto model~\cite{paso}, Kuramoto model
with time-delayed interactions~\cite{Yeung}, globally
coupled noisy bistable elements also with time delayed interactions~\cite{Huber}, and
three-state coupled oscillators~\cite{Wood2}. These cases all involve the
coexistence of a synchronous oscillatory state and a quiescent stationary state,
with the corresponding hysteresis loop.  

The novelty of our model lies in
the occurence of synchrony without destabilization of the stationary state,
so that the oscillatory state may coexist with it without exhibiting hysteresis.
In this scenario transitions between the synchronous phase and the asynchronous phase can
only be induced by a strong perturbation. 

For a purely Markovian system the master equation is an ordinary
differential equation. For coupled two-state systems it is a first order
equation, which only has a fixed point.
Therefore, in order to observe global synchronization in 
purely Markovian networks it is necessary to go beyond a two-state model~\cite{Wood,Wood2}.
Converseley, to observe synchronization in a two-state system it is necessary to go beyond a
simple Markovian model of coupled units.
The synchronization of two-state units was perhaps first observed in the
context of collective molecular motors, where spatial elastic coupling to a common environment
leads to the occurrence of oscillations~\cite{Julicher}. In globally coupled networks
of two-state units, memory effects can also lead to synchronization~\cite{Kouvaris}.
In recent studies of coupled assemblies of genetic
oscillators~\cite{Tsimring}, memory (delay) effects have been found to
play a significant role in the description of the observed synchronous dynamics 
of the on and the off (active and inactive) states of the system.

Time delays in network dynamics can be introduced in many different ways.  One is to
introduce time delays in the interaction between the elements of the
network~\cite{Yeung,Huber}. Another is to include a refractory period in the
internal dynamics of the units~\cite{Prager}. Both of these together have also been
considered~\cite{Kouvaris}. Most of the models with a time delay
that have been reported in the literature consider uniform fixed time delays, be it
in the internal dynamics and/or in the interactions.
Huber and Tsimring suggested that temporally nonuniform (that is, distributed) time
delays do not qualitatively affect the synchronization features observed in their
model~\cite{Huber}. A more detailed analysis of the role of nonuniform time delays in
synchronization has been reported in a cellular automaton version of
the susceptible-infected-recovered-susceptible model~\cite{Rozenblit}, where
a probabilistic refractory period is studied.

In this work we also consider the effects of delay times, but of an altogether different kind than those
described above. We consider identical two-state stochastic elements with a fixed refractory
period. The units are then coupled together all to all.
The interactions among the elements affect the internal dynamics in that the refractory
period becomes dependent on the time-dependent state of the entire network.  
By combining numerical and analytical 
results, we find that a global dynamics may emerge as the coupling strength is
increased beyond a critical value. We observe that synchronization occurs for some
initial conditions but not for others. Unlike the usual case of a Hopf bifurcation, the synchronized
behavior here emerges as a result of a 
saddle node bifurcation in the master equation phase space. 

The paper is organized as follows: In Sec.~\ref{sec2} we introduce our general model and
show that no Hopf bifurcation is possible for this type of dynamics. In
Sec.~\ref{sec3} we choose a specific model and
show via numerical results that a global synchronized dynamics exists for our model.
We also check the robustness of our results by considering some variants of the model.
By computing a Poincar\'{e}-type of map for the oscillatory state,
in Sec.~\ref{sec4} we demostrate that synchrony appears via a saddle node bifurcation,
and compute the critical point using this map.  We confirm the critical point by direct integration
of the generalized master equation.  We end with some concluding remarks in Sec.~\ref{sec5}.  

\section{The General Model}
\label{sec2}

\subsection{The basic unit}

\begin{figure}[]
\begin{center}
\includegraphics[width =2.7in]{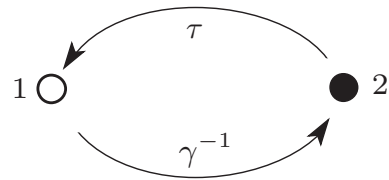}
\end{center}
\caption{Isolated binary unit. Transitions from state $1$ to state $2$ occur at
random times $\gamma^{-1}$ according to a Markov process of rate $\gamma$, while transitions
from $2$ to $1$ occur at exactly time $\tau$ after arrival at $2$. 
}
\label{fig1}
\end{figure}

Our starting point is a binary unit characterized by two states, $1$ and $2$.
Transitions from state $1$ to state $2$ are governed by a Markovian (memoryless) process
of rate $\gamma$, while transitions from $2$ back to $1$ occur after a \emph{fixed delay time} $\tau$
measured from the time of the transition to $2$ (Fig.~\ref{fig1}). 
The waiting time distribution $\psi_{1\to 2}(t)$ for a transition from $1$ to $2$ is thus an
exponential, $\psi_{1\to 2} = \gamma e^{-\gamma t}$, while that of a transition from $2$ to $1$ is a
$\delta$-function, $\psi_{2\to 1} = \delta(t-\tau)$.

We next construct the generalized master equation for the 
probabilities $P_i (t)$ to be in state $i=1,2$ at time $t$. Normalization reduces this to a single
probability, since $P _{1}(t) = 1-P _{2}(t)$.  Further, it is evident that
\begin{equation}
P _{2}(t) = \int_{t-\tau}^{t}dt' J_{1\rightarrow 2}\left(t'\right),
\end{equation}
where $J_{1\rightarrow 2}\left(t'\right)dt$ is the probability of jumping from state $1$
to state $2$ at some time within the interval $[t',t' + dt']$, with $t'$ lying in the range
$[t-\tau,t]$. (If $t'$ were to lie below this range, the jump from state $2$ back to state $1$
would have occurred already before time $t$.) In turn, $J_{1\rightarrow 2}\left(t'\right)dt'=
\gamma P _{1}(t')dt'$, the probability of being in state $1$ times the jumping rate to state $2$. 
It follows that
\begin{equation}
P _{2}(t) = \gamma\int_{t-\tau}^{t}dt' \left[1-P _{2}(t')\right].
\label{Master01}
\end{equation}
This linear integral equation for $P_2$ can also be written in differential form to arrive at the
master equation 
\begin{equation}
\mathcal{L}P _{2} = \gamma\tau,
\label{mm}
\end{equation} 
where the linear operator $\mathcal{L}$ involves infinite derivatives,
\begin{equation}
\mathcal{L} = 1 +
\gamma\tau\sum_{n=0}^{\infty}\frac{\left(-\tau\right)^n}{\left(n+1\right)!}\left(\frac{d}{dt}\right)^n.
\end{equation} 

The master equation (\ref{mm}) predicts that, after a transient, the system reaches a
stationary equilibrium, $P _{2}(t) \rightarrow P _{2}^{*}$, which is determined by 
the rate $\gamma$ and the refractory period $\tau$ via the equation
\begin{equation}
P _{2}^{*}=  \frac{\gamma\tau}{1+\gamma\tau}.
\label{Stationary1}
\end{equation} 
This result is easily understood: at equilibrium, the mean time spent by the unit in state $1$
is $\bar{\tau}_1=\gamma^{-1}$, while the mean time spent in state $2$ is exactly the
refractory period, $\bar{\tau}_2=\tau$.
The equilibrium probability $P_2^{*}$ is then simply the fraction of the time spent in state $2$,
\begin{equation}
P _{2}^{*}=  \frac{\bar{\tau}_2}{\bar{\tau}_1+\bar{\tau}_2},
\end{equation} 
which immediately leads to Eq.~(\ref{Stationary1}).

\begin{figure}[]
\begin{center}
\includegraphics[width =2.7in]{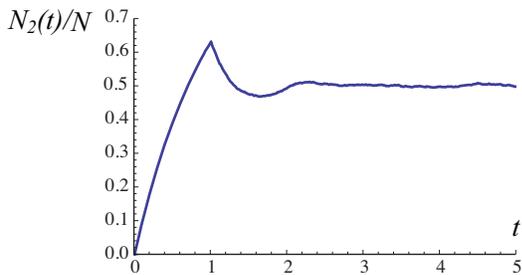}
\end{center}
\caption{Ensemble of $N=2\times 10^{4}$ isolated units with $\gamma=1$ and $\tau = 1$, which
leads to $P_{2}^{*}=1/2$. 
} \label{fig2}
\end{figure}

We can furthermore see, as follows, that the equilibrium probability $P_2^*$ is always the single
stable solution.
Equation~(\ref{Master01}) is linear and invariant under time translation. We can therefore
pick any time $t_0$ and we know that if a unit is in state $2$, it must have transitioned
there at some time during the time interval $(t_0-\tau,t_0)$. Had it transitioned to
state $2$ earlier than that, say in the interval $(t_0-2\tau,t_0-\tau)$, it would have
returned to state $1$ before time $t_0$. If it transitioned to $2$ before time $t_0-2\tau$,
it might return and be back in state $2$ at time $t_0$, but this event is already
accounted for in the accounting of transitions in the interval $(t_0-\tau, t_0)$.
One can thus write the transient solution as 
\begin{equation}
P _{2}(t) =  P _{2}^{*} + \sum_{n=1}^{\infty}q_n e^{\lambda_n t},
\end{equation} 
where the set $\left\{q_n\right\}_{n=1}^{\infty}$ depends on the ``initial condition," that is,
on the function $P _{2}(t)$ in the time interval $[t_0 -\tau,t_0]$,
and $\left\{\lambda_n\right\}_{n=1}^{\infty}$ is the set of solutions of the equation
\begin{equation}
\frac{e^{-\lambda_n\tau} -1}{\lambda_n} =\gamma^{-1}.
\label{LAM0}
\end{equation} 
The real and imaginary parts separately, 
$\lambda_n=  \rho_n + i\omega_n$, obey the equations
\begin{eqnarray}
&&e^{-\rho_n\tau} \cos\left(\omega_n\tau\right) - 1=\gamma^{-1}\rho_n,
\nonumber \\
&&e^{-\rho_n\tau} \sin\left(\omega_n\tau\right)=-\gamma^{-1}\omega_n.
\end{eqnarray}
From Eq.~(\ref{LAM0}) it is clear that the solution $\rho_n=\omega_n=0$ is spurious.
Furthermore, the choice $\rho_n >0$ leads to an immediate contradiction, namely, that $\gamma <0$.
This is impossible since $\gamma$ is a rate, hence positive.  Therefore the real parts of
the $\lambda_n$ must all be negative, and thus $P_2(t)$ goes to $P_2^*$ for any initial
condition. Indeed, the dynamics of an ensemble of $N$ such isolated units consists of a
transient of damped oscillations and eventual arrival at equilibrium, with fluctuations
due to the finite number of units. A typical evolution of $N$ such independent units is shown in
Fig.~\ref{fig2}, where we show $P_2(t) = N_2(t)/N$ vs $t$. Here $N_2(t)$ is the number of
units in state $2$. The small fluctuations just visible in the figure can be further mitigated by
working with a larger number of units.

\subsection{Coupled array: Generalized master equation}

Next we globally couple $N$ of these two-state oscillators, that is, each oscillator is
coupled to all the other oscillators. We point with special emphasis to the nature of
the coupling about to be described, because it is this feature that distinguishes
our model from others in the literature.

We take the transition rate from state $1$ to state $2$, and the refractory period
in state $2$ to become functions of the global state of the system. These are therefore
themselves now time dependent.  That is, if at a time $t$ there are $N_{1}$ units
in state $1$ and $N_{2}$ units in state $2$ (always with the constraint $N=N_{1} + N_{2}$),
the time-dependent transition
rate and refractory period are modified from their values in the isolated units,
\begin{subequations}
\begin{eqnarray}
\gamma &\rightarrow& \gamma(t) =\gamma\left(\frac{N_{2}(t)}{N}\right) \rightarrow \gamma(P_2(t)),
\label{timedep1}
\\
\tau &\rightarrow& \tau(t) =\tau\left(\frac{N_{2}(t)}{N}\right) \rightarrow \tau(P_2(t)).
\label{timedep2}
\end{eqnarray}
\end{subequations}
Here we have indicated the implementation of the thermodynamic limit $N\to\infty$,
\begin{equation}
\frac{N_{2}(t)}{N} \rightarrow P _{2}(t),
\end{equation}
and we recall again that $P _{1}(t)=1-P _{2}(t)$.
We have not yet specified the explicit form of the dependences on the state, but only that the
transition rate from $1$ to $2$ and the transition delay from $2$ back to $1$ are now functions of
the state.  Recall that in the single units, an oscillator that arrived at state $2$ at time
$t-\tau$ jumped back to state $1$ at time $t$.  In an uncoupled collection of single units, the
times at which different units are observed to leave state $2$ are random because the arrival times
are random.  However, the length of time spent in state $2$ after each arrival is fixed at the value
$\tau$. In a coupled array once again units arrive at state $2$ at random times but remain there 
for a fixed time period. Because of the random arrival, the departures are also randomly
distributed. However, there are now two differences.  The rate at which units leave state $1$ is not
fixed but depends on the instantaneous state of the coupled system.  Similarly, the waiting time in
state $2$ varies with system state.
We clarify what we mean by the time dependence of the refractory period,
without yet fixing the form of the dependence implicit in Eq.~(\ref{timedep2}): 
\emph{all the oscillators that arrived at state $2$ at time} $t-\tau(t)$ \emph{jump
back to state $1$ at time} $t$. Thus if one has a number of units in state $2$ at time $t$, some
will jump back to state $1$ right then and some not, depending on their arrival time.  We emphasize that
this is \emph{not} a stochastic feature of $\tau(t)$ but rather a consequence of the stochasticity
in the arrival times embodied in $\gamma(t)$. 

To accommodate these time dependences in a mean field description appropriate for all-to-all
coupling, we generalize the master equation (\ref{Master01}),
\begin{equation}
P _{2}(t) = \int_{-\infty}^{t}dt' F\left(P _{2}(t') \right) \Theta (t ,t').
\label{MFmodel1}
\end{equation}
Here
\begin{equation}
F\left(P _{2}(t')\right)dt' = \gamma \left(P_{2}(t')\right)\left[1 - P_{2}(t')\right]dt'
\end{equation}
is the probability to jump from state $1$ to state $2$ at some time within the interval
$[t',t' + dt']$; the factor $1-P_2(t') = P_1(t')$ is the probability to be in state $1$ at time
$t'$, and $\gamma\left(P_2(t')\right)$ is the (ever-changing) rate at which the jump
from $1$ to $2$ takes place at time $t'$. 
\emph{The factor} $\Theta(t,t')$ \emph{keeps track of this arrival rate and ``freezes" each
oscillator in state 2 for a precise length of time (refractory period) determined by the state of
the system when the oscillator arrived there}. 

We have highlighted this last term because the core and novelty of our model lie in it.
Consider the units that arrive at state $2$ at time $t'$. The number of arrivals is stochastic since
the departure from state $1$ is random.  Furthermore, the rate of random arrivals depends on the
state of the system at that time, $\gamma\left(P_2(t')\right)$. The departure of these units from
state $2$ occurs at a definite delay time $t''$ later, $t''-t' = \tau(t'')$, which depends on the
state $P_2(t'')$ at the time of departure. If $t''-t'<\tau(t'')$ we know that the units that
arrived at $t'$ have not departed.  On the other hand, if $t''-t'>\tau(t'')$ we know that they have.
Thus, if the delay time $\tau(t'')$ is not reached
at any time $t''$ in the interval
$[t',t]$, we have that $\Theta\left(t ,t'\right)=1$, an indication that all the oscillators that
arrived in state $2$ at time $t'$ are still there (while others may continue to arrive).  If the
delay time is reached at any
time within the interval, then $\Theta(t,t')=0$ because the units that arrived in state $2$ at time
$t'$ have left again. This step-like function character arises
from the fact that the delay time is fixed by the population probabilities, that is, it is not a
random variable. Explicitly, $\Theta$ is the step-like function
\begin{equation}
\Theta(t ,t')=
 \left\{
\begin{array}
[c]{ll}%
1  & \text{if  } \forall t'' \in \left[t',t\right]\text{,   } t''-t'< \tau\left[P _{2}\left(t''\right) \right] \\
0 &   \text{otherwise} \\
\end{array}
\right. .\label{Theta}
\end{equation}

We call the condition that leads to $\Theta(t,t')=1$ \emph{the inequality}.  In Appendix~\ref{app} we
explore this question in detail independently of the particular functional form of $\gamma$ and
$\tau$ as functions of time. Here we take advantage of the lessons learned there and implement the
behaviors that arise in our case. 

Our ultimate goal is to find the behavior of the $P_i(t)$ at long times.  In particular, we are
interested in establishing whether this behavior is time dependent or time independent, and 
under what conditions one might observe the former or the latter.

\subsection{Stationary solutions}
\label{stationary}

We start by exploring the possible occurrence of one or more stationary states, that is, 
one or more time-independent solutions $P _{2}(t) = P _{2}^{*}$.
This case is covered by the first part of Appendix~\ref{app} and falls in the class of behaviors
described by the generalized master equation~(\ref{MFmodel3}),
\begin{equation}
P _{2}(t)= \int_{t-\tau(P_2(t))}^{t}dt' F\left(P _{2}(t') \right).
\label{MFmodel3text}
\end{equation}
Specifically, such a stationary solution would satisfy
\begin{equation}
P _{2}^{*}=  \frac{\gamma\left(P _{2}^{*}\right)\tau\left(P _{2}^{*}\right)}{1+\gamma\left(P _{2}^{*}\right)\tau\left(P _{2}^{*}\right)}.\label{Stationary2}
\end{equation} 
Note that this is only similar to Eq.~(\ref{Stationary1}) for the uncoupled system in aspect since
Eq.~(\ref{Stationary2}) is in general a non-linear equation for 
quiescent states. This nonlinear equation could have more than one solution
(multi-stability) or even no solution at
all, depending on the particular functional forms chosen for
$\gamma\left(P _{2}\right)$ and $\tau\left(P _{2}\right)$.

\subsection{Absence of Hopf-type instabilities}

While we have not yet established whether Eq.~(\ref{Stationary2}) actually has one or
more solutions,
we can show that the coupled array does not support Hopf-type instabilities leading to oscillatory
or chaotic solutions.  Indeed, a \emph{linear} analysis at best leads
to other quiescent solutions. To show this we introduce a perturbation,
\begin{equation}
P _{2}(t)=  P _{2}^{*} + \varepsilon e^{\mu t},
\end{equation} 
with $\left|\varepsilon\right|\ll 1$. First, we note that
if $\tau'(P_2(t)) \sim \mathcal{O}(1)$, as it will be in our model explicitly introduced later,
then $d\tau\left(P _{2}(t)\right)/dt \sim \varepsilon <1$.
(The prime denotes a derivative with respect to the argument.)
According to the discussion in the appendix, the generalized master equation~(\ref{MFmodel3text})
is valid if $\mu$ is negative and for sufficiently short times even if $\mu$ is positive. To first
order in $\varepsilon$, the
master equation is then given by
\begin{align}
\varepsilon e^{\mu t}=& \varepsilon\int_{t-\tau\left(P _{2}^{*}\right)}^{t}dt' F'\left(P _{2}^{*}
\right)e^{\mu t'}
\nonumber \\
&+ \int_{t-\tau\left(P _{2}^{*}\right)-\varepsilon\tau'\left(P _{2}^{*}\right)e^{\mu t}}^{t-\tau\left(P _{2}^{*}\right)}dt' F\left(P _{2}^{*} \right)
\nonumber.
\end{align}
Straight substitution of Eq.~(\ref{Stationary2}) in this equation shows that $\mu$ must satisfy
\begin{equation}
\frac{e^{-\mu\tau^{*}} -1}{\mu} =\Gamma^{-1},\label{LAM}
\end{equation} 
where
\begin{equation}
\tau^{*}\equiv \tau\left(P _{2}^{*}\right), \qquad
\Gamma=  \frac{F'\left(P _{2}^{*} \right)}{\tau'\left(P _{2}^{*}\right)F\left(P _{2}^{*} \right)
-1}.
\nonumber
\end{equation}
Therefore, all the information about the linear stability of $P _{2}^{*}$
resides in the two numbers $\tau^{*}$ and $\Gamma$. No additional information about
the specific functional form of $\tau(P_2)$ or of $\gamma(P_2)$ is required. 

To analyze equation (\ref{LAM}) one can separate the real and imaginary parts,
$
\mu=  \beta + i\Omega,
$
in terms of which condition~(\ref{LAM}) can be separated into the two equations 
\begin{align}
1 - e^{-\beta\tau^{*}} \cos\left(\Omega\tau^{*}\right)&=\Gamma^{-1}\beta,
\nonumber \\
e^{-\beta\tau^{*}} \sin\left(\Omega\tau^{*}\right)&=\Gamma^{-1}\Omega
\nonumber.
\end{align}
A Hopf-type instability requires that a stable stationary solution
become unstable with non-zero frequency. A critical point must satisfy 
$
\beta_c=0 \text{  with  } \Omega_c\neq0, 
$
that is,
\begin{equation*}
\cos\left(\Omega_c\tau^{*}\right)=1 \text{  and  } \sin\left(\Omega_c\tau^{*}\right)
=\Gamma^{-1}\Omega_c.
\end{equation*}
However, this pair of equations only has the solution $\Omega_c=0$. Note that the
solution $\mu=0$ is no longer necessarily spurious as it was in the decoupled array.  There the
solution $\lambda_n=0$ required that $\gamma\tau=-1$ [cf. Eq.~(\ref{LAM0})],
which is impossible because both parameters are positive.  However, here the condition
$\Gamma\tau^{*}=-1$ can be seen as an equation for a critical value of the control parameter.

In summary, this linear analysis predicts no complexity beyond possible
bifurcations between stationary solutions.
However, the analysis itself is limited:
our nonlinear infinite-dimensional system has the potential to exhibit more complex
behavior such as oscillatory or chaotic orbits.  We will next explore the possible occurrence
of limit cycles (synchronization of the units) via global bifurcations, that is, bifurcations
that go beyond the vicinity of a fixed point to
involve larger regions of phase space. For low dimensional systems there are three generic
global bifurcation pathways to cycles~\cite{StrogatzLib}: saddle-node, infinite-period
and homoclinic bifurcations.  We find below that our infitine-dimensional system also
follows one of these pathways.

\subsection{Generic features beyond stationary states}

We continue to explore possible behaviors, now away from stationary states. Our analysis in
Appendix~\ref{app} shows that there is only a limited array of possible behaviors.  We will
basically encounter only two.

The first, which we will call behavior $B$, occurs when $\Theta(t,t')$ can be expressed as a 
Heaviside function, cf. Eq.~(\ref{steadystate}) in the appendix,
\begin{equation}
\Theta(t,t')=\theta \left(t' - t + \tau(P_2(t)) \right). 
\label{regimenB}
\end{equation} 
As shown in the appendix, this expression is valid when (1) $\tau(t)$ is continuous and
piecewise differentiable; (2) where differentiable, $d\tau/dt <1$; (3) $\tau(t)$ may have
discontinuities presenting a \emph{downward} jump from a high value to a low value as time
increases, but not from a low value to a high value.  

\begin{figure}[]
\begin{center}
\includegraphics[width =2.7in]{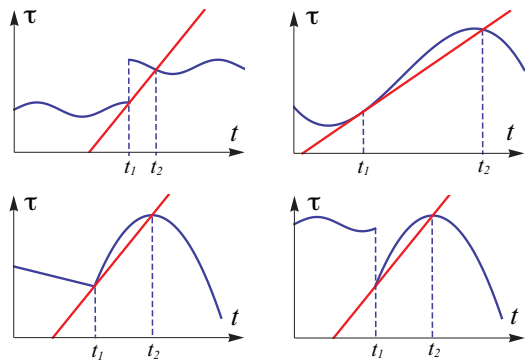}
\end{center}
\caption{Four ways to enter regime A from regime B.  Upper left: when there is
a discontinuity in $\tau$ from a low value to a high value. Upper right: when $d\tau/dt$ smoothly
goes through the value $d\tau/dt=1$ from whence it increases. Lower left: similar to upper right
except that $\tau(t)$ need not be differentiable at the point of entry in A. Lower right: here
there is a discontinuity in $\tau$ from a low value to a high value together with a derivative that
approaching the point of discontinuity that exceeds unity, $\lim_{t\to t_1^+}(d\tau/dt) >1$.
In all four cases entry into regime A occurs at time $t_1$ and exit at time $t_2$.
}
\label{schematic3}
\end{figure}

The other behavior we will encounter, behavior A, occurs in one of two ways.  It can come about
because (3) above is violated, that is, $\tau(t)$ exhibits a discontinuity presenting an
\emph{upward} jump from a low value to a high value, or because condition (2) above is violated.
There are several ways to violate the condition $d\tau/dt <1$ and thus to enter regime A. These
four ways are illustrated in Fig.~\ref{schematic3}. One, illustrated in the upper right panel of the
figure, occurs smoothly. It starts at time $t_1$ where $d\tau/dt|_{t=t_1}=1$.
A second way to enter regime A is illustrated in the lower left panel.  Here regime A again begins
at time $t_1$, but $\tau(t)$ need not be differentiable at $t_1$.  Instead, here
$\lim_{t\to t_1^+}(d\tau/dt) >1$.
The third way to enter regime A, illustrated in the lower right panel,
is via a discontinuity from a high value to a low value at time $t_1$, but with a violation of
condition (2) above, that is, again $\lim_{t\to t_1^+}(d\tau/dt) >1$.
The final way, illustrated in the upper left panel, occurs when there is a discontinuity
from a low value to a high value. 
In all four cases to end regime A and re-enter behavior B it is not sufficient for $d\tau/dt
<1$ for some $t>t_1$.  The end point occurs when $\tau(t)$ crosses the line $t-t_1+\tau(t_1)$,
that is, at time $t_2$ that satisfies 
\begin{equation}
\tau(t_2) = t_2 - t_1 + \tau(t_1).
\end{equation}
The associated $\Theta$ then is
\begin{equation}
\Theta(t,t') = \theta(t'-t_1+\tau(t_1)), \quad t\in [t_1,t_2].
\label{forA}
\end{equation}

We conclude with a comment: we have indicated that in all cases regime $A$ ends with a re-entry into
Regime $B$ in a finite time.  This is, in fact, not necessary.  Regime $A$ could continue forever.
But this would imply a continual growth of $\tau(t)$, which leads to the uninteresting conclusion
that (since exit from state $2$ of our array units is forever slowed down) all units will end up in
state $2$ and will remain there. This is a trivial dynamics that will not be pursued further.

We end this section by noting that we have so far made no specific assumptions about the functional
forms of $\gamma(P_2(t))$ or $\tau(P_2(t))$ but have, instead, discussed the problem in full
generality.  We also note that our discussion so far is fully analytic.  Now, to proceed we
introduce specific forms for these functional dependences.  

\section{The specific model: synchronization}
\label{sec3}

\subsection{Choosing a model}

At this point any further analysis of the emergence of synchronization in our ensemble of
globally coupled oscillators requires us to specify 
the time-dependent quantities $\gamma(t)$ and $\tau(t)$. For the rate $\gamma$ we choose
the standard (Arrhenius-type) exponential law~\cite{Wood,Wood2,Prager,Kouvaris}
\begin{eqnarray}
\gamma(t)&=& g\exp\left(\frac{a[N_{2}(t)-N_1(t)]}{N}\right)
\nonumber\\
&=& g\exp\left(\frac{a[2N_{2}(t)-N]}{N}\right).
\label{CouplingGamma} 
\end{eqnarray}
For the refractory period we introduce a non-monotonic dependence on $N_{2}(t)/N$, 
\begin{equation}
\tau(t) =  \tau_{0}\frac{N_{2}(t)N_1(t)}{N^{2}} =  \tau_{0}\frac{N_{2}(t)(N-N_{2}(t))}{N^{2}},
\label{tau2}
\end{equation}
which exhibits a maximum when half the units are in state $1$ and the other half in state $2$.
An imbalance in either direction shortens the refractory period.
Here $g$, $a$ and $\tau_0$ are coupling parameters.
Next, we can scale time so that with no loss of generality one of the parameters can be fixed,
say $g=1$, and hence the system has a total of three free
parameters: $a$, $\tau_{0}$ and $N$. We focus on the thermodynamic behavior $N\gg 1$,
so that the phase space diagram of the system is, in fact, the two-dimensional
parameter plane $\left(a,\tau_{0}\right)$. We have observed that varying
just one of these two parameters allows us to capture the main features of
the system (the transition to synchronization), so we fix $\tau_{0}=2$ 
and present our analysis as a function of the coupling parameter $a$.

Note that in principle the coupling parameter may be positive or negative.  A positive parameter
implies that if many units are in state $1$, say, then they leave that state more slowly than if
there are only a few.  We choose to study the case of negative $a$, where the behavior is the
opposite, namely that units leave state $1$ at a greater rate when it is highly occupied.  That is,
the system ``avoids" crowding.  This is consistent with the form chosen for the refractory period:
it, is shortened when either state is overcrowded.

\subsection{Outcomes in the thermodynamic limit}

\begin{figure}[]
\begin{center}
\includegraphics[width =3.1in]{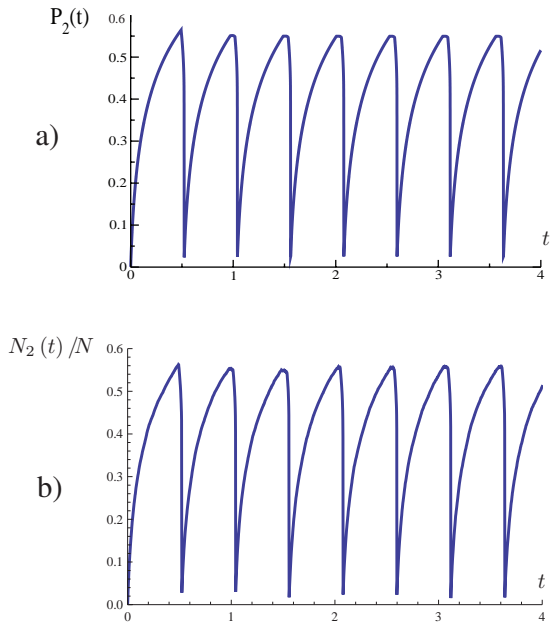}
\end{center}
\caption{(Color online) Long-time behavior of the
model for $a=-2$ and $\tau_{0}=2$.
(a) Numerical solution of the mean field equation (\ref{MFmodel1}). 
(b) Direct numerical simulations of the equations of motion for
$N=10^{4}$ units coupled according to the prescriptions (\ref{CouplingGamma}) and
(\ref{tau2}).
}
\label{fig4}
\end{figure}

First we test the mean field theory for our model, that is, we solve Eq.~(\ref{MFmodel1}), which of
course must be done numerically.  We observe that after a transient dynamics, the system approaches
a long-time behavior that depends on
the choice of the coupling parameter $a$.
When $|a|$ is small (``small" to be specified accurately
later), the system approaches a quiescent state well described by Eq.~(\ref{Stationary2}). On the
other hand, when the coupling is sufficiently strong (to be determined later),
the system either approaches this quiescent state, or,
following a transient, the system settles into a globally synchronized dynamics wherein the
probabilities $P_i(t)$  oscillate in highly anharmonic but essentially periodic fashion.
Fig.~\ref{fig4}(a) displays a typical such global limit cycle.  We thus observe the coexistence
of a stationary state and a stable limit cycle.
Note that we can not necessarily simply search for initial conditions $P_2(0)$ that lead to one behavior
or the other.  Because the system has a memory, the ``basins of attraction" in general would have to be
characterized in functional space and require knowledge of $P_2(t)$ for past times before $t=0$.
For our model, we need to know $P_2(t)$ at most in the interval $[-\tau_0/4,0]$. If all the units
are initially in state $1$, that is, if $P_2(0)=0$, then we do not need information about the past.
We have not attempted to unravel the functional initial condition problem here.

We test the mean field master equation by also directly simulating the equations of motion of
our system.  Our simulations here are carried out for $N=10^4$.
Here again we find that for sufficiently strong coupling,
after a transient dynamics the system approaches one of two
long-time behaviors: the 
system either converges to a stationary state wherein $N_1$ and $N_2$ approach essentially
constant values (again well described by the mean field quiescent state~(\ref{Stationary2})),
or the system goes to a final state of synchronized dynamics that consists of
coherent oscillations of $N_{1}$ and $N_{2}$. The number $N=10^4$ is sufficiently large for us not
to observe fluctuations that would perturb the system out of an oscillatory dynamical behavior
during the time of our simulations, but presumably a sufficiently long run would exhibit such
transitions.  In any case, we thus again observe the coexistence of a
stationary state and a stable limit cycle.
Fig.~\ref{fig4}(b) displays a typical such 
global limit cycle, this one for the same initial condition and coupling as in Fig.~\ref{fig4}(a).
We see that $N_{2}\left(t\right)/N$ exhibits very anharmonic oscillations
with well defined amplitude and frequency.  The similarity between the
two panels reassures us that $N=10^4$ is sufficiently large to 
capture all the essential features of the thermodynamic behavior of the system.
The excellent agreement between the solution of the
mean field master equation and the direct numerical simulations is gratifying. The simulations
inevitably exhibit very small fluctuations not present in the mean field model because the
simulations are carried out for finite, albeit large, $N$.

We will subsequently argue that the two solutions, synchronous and asynchronous, are co-existing
stable solutions, and that the synchronization is here not related to the destabilization of the
stationary state. Regardless of initial condition, a sufficiently strong external disturbance
applied to the system can drive it
from one state to the other. In a system with a finite number of units the two solutions are metastable states not impervious to fluctuations; indeed, a
sufficiently strong fluctuation can cause a transition from one of these states to the other. Thus, now the perturbation need not be external to the system.

In the next subsection we present a number of results that arise from
direct numerical simulations of coupled units.
Then, in the next section, we
return to the master equation and, to understand the synchronization properties from this more
analytic point of view, we construct a Poincar\'e map for the oscillatory state.
Among other results, we calculate the critical coupling for our model in the thermodynamic limit.

\subsection{A variety of results of direct numerical simulations}

We present several sets of results, not all meant to be quantitatively accurate but rather to
illustrate a number of interesting observed features of our system. More detailed studies of these
features will be explored in subsequent work~\cite{subsequent}.

\begin{figure}[h!]
\begin{center}
\includegraphics[width =2.9in]{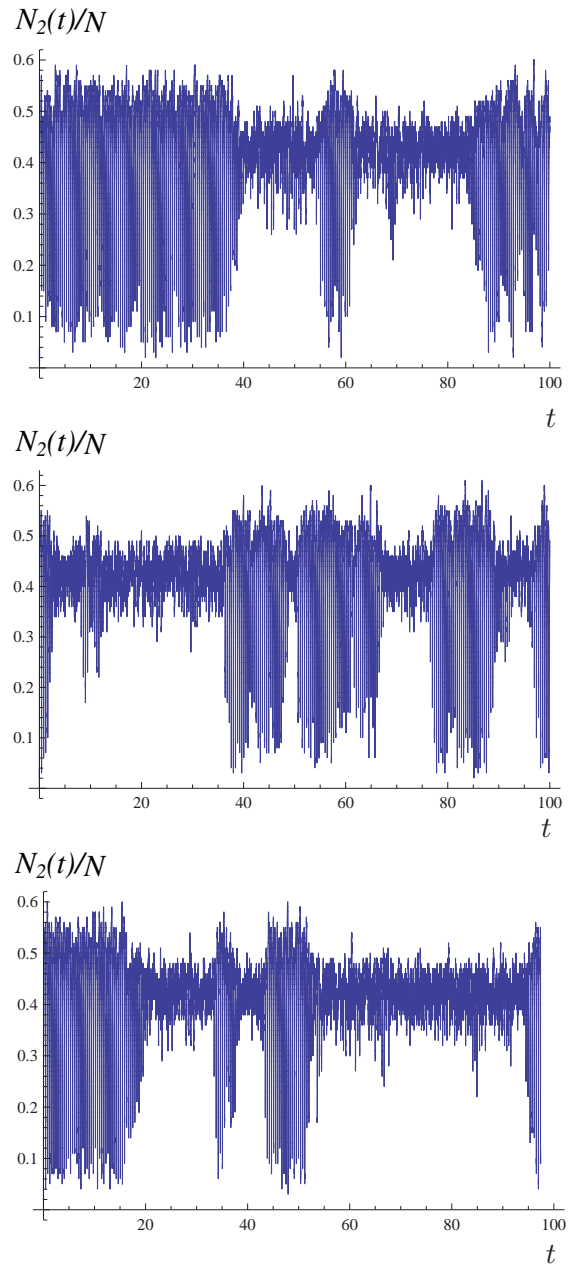}
\end{center}
\caption{(Color online) Typical realizations for an array of $100$ units
coupled according to the prescriptions (\ref{CouplingGamma}) and (\ref{tau2})
above critical coupling.
The parameter values are the same for the three realizations.  The mean field stationary value for
these parameters is $P_2^*\simeq 0.422$.
}
\label{old3}
\end{figure}

First, we illustrate the fact that for small $N$ the fluctuations are sufficiently strong to
drive the system back and forth
between oscillatory and stationary behavior.
We have not explored the dependence of the rate of these transitions on $N$, but note that
they arise from the microscopic dynamics.
Three realizations of such a progression in time are shown in Fig.~\ref{old3} for an array
of $N=100$ all-to-all coupled
units. The precise parameter values are not particularly important other than to confirm that the
stationary episodes in the figure are consistent with the stationary value obtained from
Eq.~(\ref{Stationary2}), which here is $P_2^* \simeq 0.422$.

A second point to explore is the robustness of the results to variations in the model.  Ours would not
be a very interesting effort if it depended strongly on the specific features of the model without
room for variation.  We have explored two specific variations. 

\begin{figure}[h!]
\begin{center}
\includegraphics[width =2.9in]{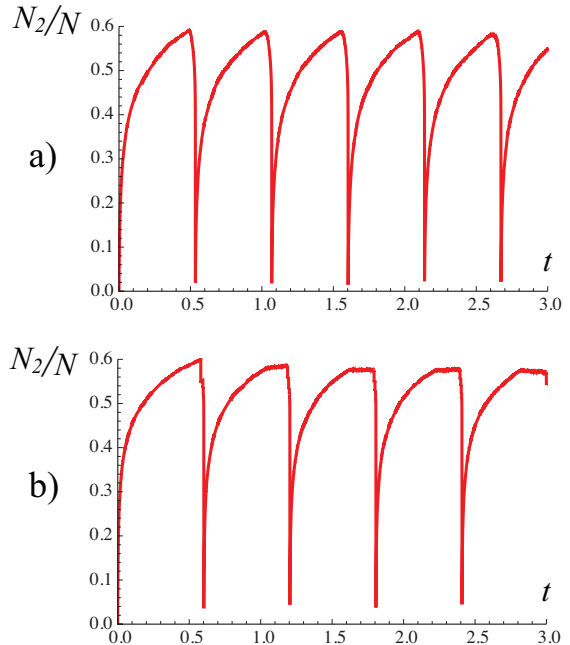}
\end{center}
\caption{(Color online) (a) Typical realization for direct simulations of the model described
in Eqs.~(\ref{RateModel2}) and (\ref{MFModel2}) with $\tau_0=2$, $a=-4$, $b=100$, and $N=10,000$
coupled according to the prescriptions (\ref{CouplingGamma}) and (\ref{tau2}). (b) Typical
realization for model with ``displaced waiting time" with parameters $\tau_0 =2$, $a=-5$,
$\delta\tau_0=0.1$, and $N=10,000$.
}
\label{othermodels}
\end{figure}

To display the first generalization we note that we can express our basic model more generally than
we have done, namely, we can introduce a rate $g_1(t,t')$ of transitions from state $1$ to state
$2$, and likewise a rate $g_2(t,t')$ for transitions from state $2$ back to state $1$.  Here $t'$
is absolute time, and $t$ denotes the waiting time.  We continue to assume that transitions from
$1$ to $2$ constitute a Markov process so that $g_1(t,t')\equiv\gamma(t')$ as
indicated in Eq.~(\ref{CouplingGamma}).  However, for the return we allow a distributed waiting
time of the form 
\begin{equation}
g_{2}\left(t, \frac{N_{2}\left(t'\right)}{N}\right) =
\frac{1}{\tau_{0}}\left(\frac{t}{\tau\left(N_{2}\left(t'\right)/N\right)}\right)^{b},\label{RateModel2}
\end{equation} 
which reduces to our working model with a single waiting time
when $b\to\infty$. The full master equation is
\begin{align}
P_2(t) = &\int_{0}^{\infty}dt'' \gamma\left(P_2(t - t'') \right)\left(1
- P_2(t - t'') \right) 
\nonumber \\
& \times\exp\left(- \int_0^{t''}\frac{dt'}{\tau_0} \left(\frac{t'}{\tau\left(P_{2}\left(t- t'' + t'\right)\right)}\right)^{b}\right),\label{MFModel2}
\end{align}
which reduces to Eq.~(\ref{MFmodel1}) with (\ref{Theta}) when $b\to\infty$. We have ascertained
for various values of $b$ that this also leads to oscillatory solutions for sufficiently strong
coupling, that these coexist with stationary solutions in this regime, and that the solutions
obtained from the simulations of a large number of coupled units and those obtained from the
generalized master equation coincide.  The stationary solutions of the master equation satisfy
\begin{equation}
P _{2}^{*}=  \frac{C\gamma\left(P _{2}^{*}\right)\left[\tau_{2}\left(P
_{2}^{*}\right)\right]^{\frac{b}{b+1}}}{1+C\gamma\left(P _{2}^{*}\right)\left[\tau_{2}\left(P
_{2}^{*}\right)\right]^{\frac{b}{b+1}}},\label{StationaryModel2}
\end{equation} 
with
\begin{equation*}
C = \left[\tau_{0}\left(b+1\right)\right]^{\frac{1}{b+1}}\Gamma\left(\frac{b+2}{b+1}\right).
\end{equation*} 
Here $\Gamma$ is the gamma function.  It is straightforward to verify that this reduces to
Eq.~(\ref{Stationary2}) when $b\to\infty$. The oscillatory solutions exhibit the same strongly
anharmonic behavior for the values of $b$ we have explored (in the neighborhood of $b=100$) as they
do in our working model. A typical realization of such a stationary solution is shown in
the upper panel of Fig.~\ref{othermodels}.

We have also noted that $P_2=0$ is a stationary solution of our working model as well as of
the generalization discussed above.  To vary this, we have explored the consequence of modifying the
waiting time in state $2$ from the form $\tau(t)$ of Eq.~(\ref{tau2}) to $\tau(t) + \delta \tau_0$.
We have tested the effect of this addition for small values, $\delta \tau_0\sim 0.1$, and have
again found the same behaviors. A typical realization of an oscillatory solution for this model is
seen in the lower panel of Fig.~\ref{othermodels}.

\begin{figure}[]
\begin{center}
\includegraphics[width =3.0in]{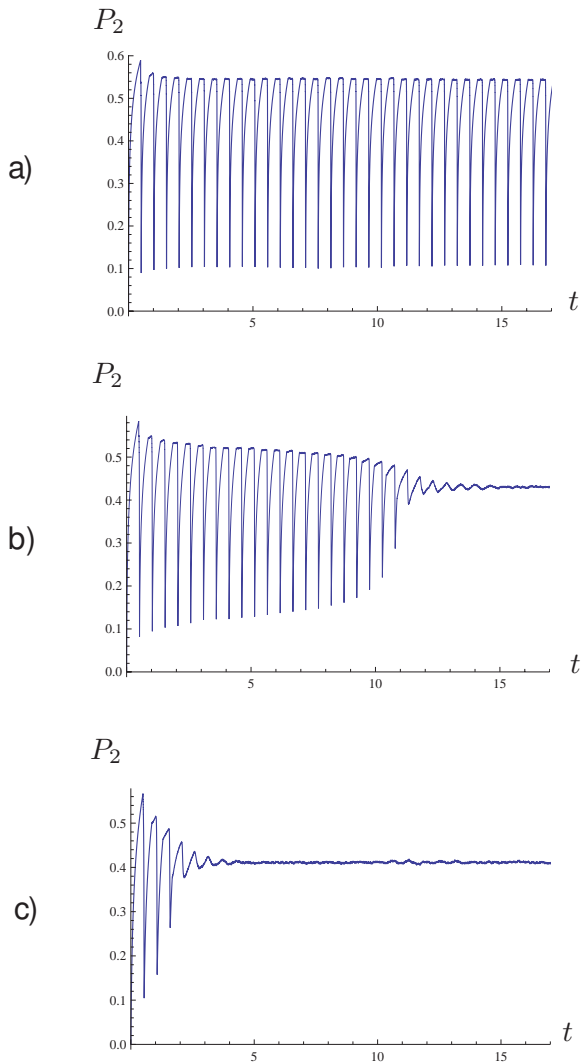}
\end{center}

\caption{Numerical simulation of $N = 160,000$ units: (a) well above critical coupling;
(b) just below critical coupling; (c) well below critical coupling.
}
 \label{figold6}
\end{figure}

Finally, we illustrate in Fig.~\ref{figold6} the loss of synchronization with our working model
as we cross the critical
coupling.  The simulation in the figure is for $160,000$ coupled units. In panel (a) the coupling is
well above the critical value; in (b) it is just below the critical value, and in (c) it is well below
this value.

As promised, we now return to the master equation and, to understand the synchronization
properties from this more analytic point of view, we construct a Poincar\'e map for the
oscillatory state so as to obtain a
quasi-analytic estimate of the critical coupling in the thermodynamic limit.

\section{Poincar\'{e}-type map for the oscillatory state}
\label{sec4}
\begin{figure}[]
\begin{center}
\includegraphics[width =3.5in]{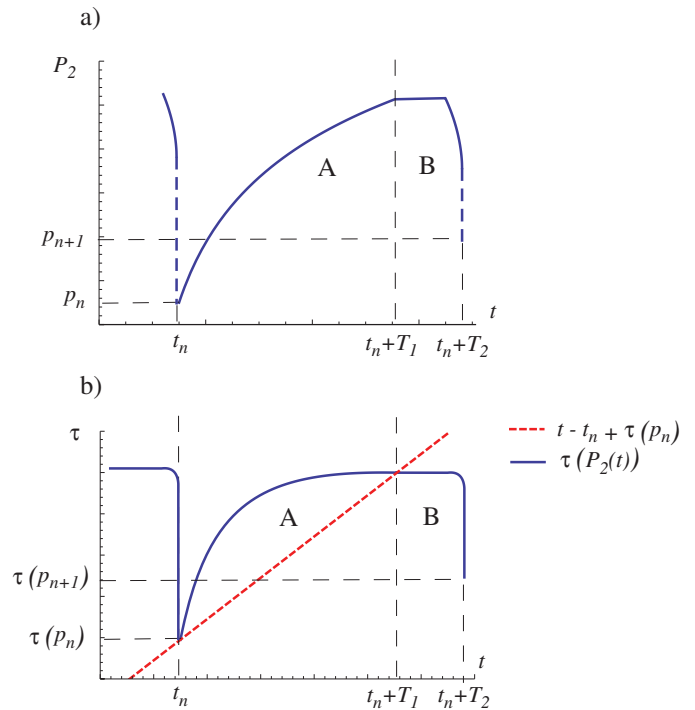}
\end{center}
\caption{(Color online) (a) $P_{2}$-profile during the $n$-th cycle.  (b) Transition delay time
$\tau$ vs time (solid, blue), and the straight line $\tau=t-t_n+\tau(p_n)$ vs time (dashed, red).
The A portions use the analytic result Eq.~(\ref{solveA}), while the results for 
regime B are outcomes of the numerical integration of Eq.~(\ref{MFmodel1B-dif}). Parameter values:
$\tau_{0}=2$, $a=-1.5$ and $p_{n} = 0.05$ ($p_{n+1}$ was inserted manually in order to clarify the
plots).} \label{fig5}
\end{figure}

To understand the synchronization properties of our coupled array, and guided by our numerical
solutions but proceeding as analytically as possible, we focus on a single
typical oscillatory cycle, cf. Fig.~\ref{fig5}(a).
We define a cycle as the evolution between the two evident discontinuities in the figure,
indicated by dashed lines. 
In more detail, we observe that at the start of a cycle, $P_2(t)$ at first grows gradually and
then it quickly decreases to a minimum value that is in general different from the
starting value of the cycle. Then the process begins again.
The cycle shown in Fig.~\ref{fig5} is a generic $n$-th cycle, where $t_n$
denotes an arbitrary time which does not play any explicit role in the calculations. The
$n$-th cycle starts from some initial condition $P_2(t_n) = p_n$ which is the end point of the
previous discontinuity (and whose value depends on the past).  After a time
$T_2$ the system undergoes another discontinuity ending at time $t_{n+1} = t_n+T_2$, that is,
$P_2(t_{n+1}) = p_{n+1}$. Note that $T_1$ and $T_2$ change from cycle to cycle because periodicity
is not perfect, that is, they both depend on $n$ via $p_n$. 
The key to the synchronization is the Poincar\'{e}-type map
\begin{equation}
p_{n+1} = f\left(p_{n}\right).
\label{MAP}
\end{equation}
The oscillatory solution corresponds to a fixed point of the map, $p^{*} =
f(p^{*})$, and it is stable if $\left|f'\left(p^{*}\right)\right|<1$. Otherwise
it is unstable~\cite{StrogatzLib}.

In order to solve the master equation (\ref{MFmodel1}) and 
compute the map (\ref{MAP}), we note that each cycle exhibits two significantly
different regimes of behavior, indicated as A and B in Fig.~\ref{fig5}.
These are precisely the regimes A and B discussed in detail in Appendix~\ref{app} and in
Sec.~\ref{sec2}.

{\bf A-regime:} This regime is characterized by the fact that the curve $\tau(P_2(t))$ 
lies above the line $t-t_n+\tau(p_n)$, 
$\tau\left(P _{2}(t) \right) > t - t_{n} + \tau(p_{n})$, for $t_n < t < t_n+T_1$. 

{\bf B-regime:} Here the inequality is reversed (the curve lies below the line), 
$\tau\left(P _{2}(t) \right) < t - t_{n} + \tau(p_{n})$, for $t_n+T_1 < t < t_n+T_2$.

We next consider the generalized master equation in each regime separately.  Consider first regime A.
From Eq.~(\ref{forA}) it follows that for $t$ within the interval $[t_n, t_n+T_1]$,
\begin{equation}
\Theta(t,t') = \theta(t'-t_n+\tau(t_n)), 
\end{equation}
where for economy of notation we have written $\tau(P_2(t_n))\equiv \tau(t_n)$.
The master equation can then be written as 
\begin{equation}
P _{2}^A(t) = \int_{t_n-\tau(t_n)}^{t}dt' F\left(P _{2}(t') \right).
\label{MFmodel1mod1}
\end{equation}
We can alternatively write this in the differential form
\begin{equation}
\dot{P} _{2}^{A} = F\left(P _{2}^{A}\right),\label{MFmodel1A}
\end{equation}
with the initial condition $P _{2}^{A}(t_{n};p_{n})=p_{n}$.
The solution is
\begin{equation}
P _{2}^{A}\left(t;p_{n}\right) = 1-
\frac{1}{2a}\text{ei}^{-1}\left(\text{ei}\left(2a\left(1-p_{n}\right)\right)-e^{
a}\left(t-t_{n}\right)\right),
\label{solveA}
\end{equation} 
where $\text{ei}$ denotes the exponential integral, 
\begin{equation*}
\text{ei}\left(z\right) = -\int_{-z}^{\infty}du
\frac{e^{-u}}{u},
\end{equation*}
and $\text{ei}^{-1}$ denotes its inverse. Hence, in this regime
we have presented an exact analytic solution.

We carry this analytic focus further by noting that
the A-regime ends after a time interval $T_{1}$, which is defined by the equation
\begin{equation}
T_{1}+ \tau\left(p_{n}\right)=\tau\left(P _{2}^{A}\left(T_{1}+t_{n};p_{n}\right)
\right),
\end{equation}
which identifies the point at which the $\tau$ curve and the line $t-t_n+\tau(p_n)$ intersect.
This identification is important because $P_2(t)$ reaches a maximum at this point (followed by a
plateau discussed below), and with the analytic solution 
~(\ref{solveA})
we can compute this
maximum and the time at which it occurs.
Note again that $T_{1}$ depends on $p_{n}$, but does not depend on $t_{n}$.

In the B-regime we are forced to rely partly but minimally on numerical simulations,
which lead to the observation that $d\tau\left(P _{2}^{B}\left(t;p_{n}\right)
\right)/dt<1$. Here we have explicitly indicated the dependence of $\tau$ on $t$ via $P_2^B$.
The master equation in this regime can be written as
\begin{equation}
P _{2}^{B}\left(t;p_{n}\right) = \int_{t - \tau\left(P
_{2}^{B}\left(t;p_{n}\right) \right)  }^{t}dt' F\left(P
_{2}\left(t';p_{n}\right) \right) 
\label{MFmodel1B}
\end{equation}
(cf. Eq.~(\ref{MFmodel3})).
Taking a derivative of this equation with respect to $t$ and solving for $\dot{P}_2^B(t;p_n)$
yields the differential form
\begin{equation}
\dot{P} _{2}^{B}(t;p_n)= \frac{F\left(P
_{2}^{B}(t,p_n)\right)-F\left(\mathcal{P}\left(t-\tau\left(P
_{2}^{B}\right)\right)\right)}{1-F\left(\mathcal{P}\left(t-\tau\left(P
_{2}^{B}\right)\right)\right)\tau'\left(P _{2}^{B}
\right)},\label{MFmodel1B-dif}
\end{equation}
where
\begin{equation*}
\mathcal{P}\left(t\right)=
 \left\{
\begin{array}
[c]{ll}%
P _{2}^{B}\left(t;p_{n-1}\right)   & \text{if }   t < t_{n} \\
P _{2}^{A}\left(t;p_{n}\right)  &   \text{if } t_{n} < t < t_{n} + T_{1}\\
P _{2}^{B}\left(t;p_{n}\right)  &   \text{if } t > t_{n} + T_{1}\\
\end{array}
\right. ,
\end{equation*}
together with the initial condition
\begin{equation}
P_{2}^{B}\left(T_{1}+t_{n};p_{n}\right)=P
_{2}^{A}\left(T_{1}+t_{n};p_{n}\right).
\end{equation}
Equation~(\ref{MFmodel1B-dif}) is a non-autonomous differential equation
for $t - \tau\left(P _{2}^{B} \left(t;p_{n}\right) \right)< t_n$
because it depends on $P_2^B$ in the previous cycle, $P_2^B(t;p_{n-1})$.
It is also non-autonomous for 
$t_n < t - \tau\left(P _{2}^{B} \left(t;p_{n}\right) \right)< t_n + T_1$
because it depends on $P_2^A$ in the same cycle, $P_2^A(t;p_n)$.
In general, for $t - \tau\left(P_{2}^{B}\left(t;p_{n}\right) \right)  > t_{n} + T_{1}$ it
is a delay differential equation~\cite{Wiener} because it only $P_2^B$ at
previous times; however, as seen immediately below, this regime is not reached in our system in the
oscillatory state. 

More specifically for our case, the description of the system in regime B then proceeds as follows.
For $t - \tau\left(P _{2}^{B}\left(t;p_{n}\right) \right)  < t_{n}$, the
$P_{2}$-profile exhibits a plateau
as seen in Fig.~\ref{fig5}(a).
Note that this is a word used here to describe a numerical 
observation rather than an analytic precise outcome.
Then, for $t_{n} < t
- \tau\left(P _{2}^{B}\left(t;p_{n}\right) \right)  < t_{n} + T_{1}$ it quickly
decreases up to the time at which it exhibits the discontinuity. This occurs at $t_{n} +
T_{2}$, which satisfies $T_{2} - \tau\left(P _{2}^{B}\left(t_{n} +
T_{2};p_{n}\right) \right)  < T_{1}$.  Equation (\ref{MFmodel1B-dif})
remains a non-autonomous differential equation during this entire portion of the cycle,
that is, it reaches the discontinuity before it becomes a delay differential equation.

At the level of Eq.~(\ref{MFmodel1B-dif}) the discontinuity is related to the
divergence $\dot{P} _{2}^{B}\left(t\rightarrow t_{n}+T_{2};p_{n}\right)
\rightarrow -\infty$.
At the discontinuity, $P_2(t)$ falls to the value $p_{n+1}$, and regime A begins again,
\begin{equation}
p_{n+1} = P_2^A(t_{n+1}; p_{n+1}).
\end{equation}
We also note that numerical simulations show that $p_n$ is small, so that we
can assume that $T_2 - \tau(t_{n+1}) > T_1$.  In this case, $t$ lies in the range,$[t_n + T_2
-\tau(t_{n+1}), t_n + T_2]$.  This places us in regime B of cycle $n$.  We can thus write
\begin{equation}
p_{n+1} = \int_{t_{n+1} -\tau(p_{n+1})} ^{t_{n+1}} dt' F\left(P_2^B (t';p_n)\right)
\label{IntMap}
\end{equation}
where $t_{n+1}=t_n+T_2$ and where we have invoked Eqs.~(\ref{IntMap}) and (\ref{MFmodel1mod1}).
This is an equation that relates $p_{n+1}$ to $p_n$, that is, it implicitly
contains the map (\ref{MAP}).
\begin{figure}
\begin{center}
\includegraphics[width =3.1in]{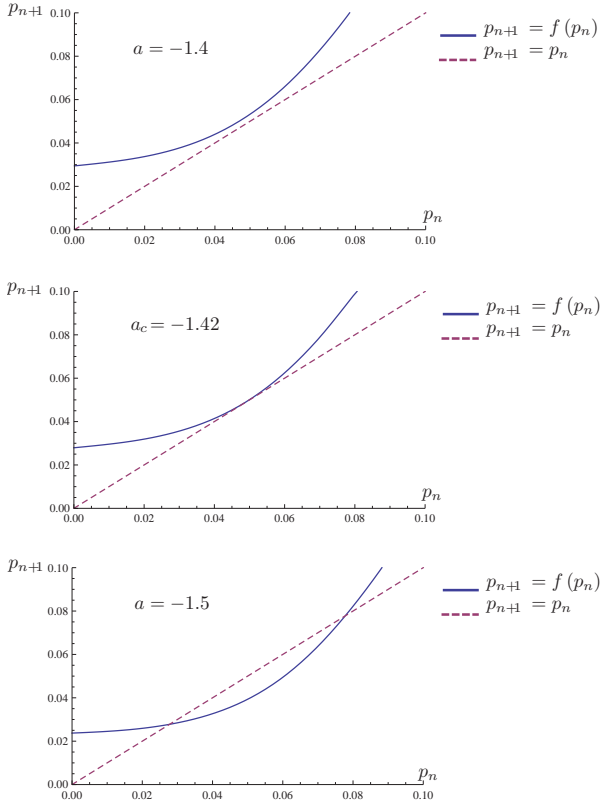}
\end{center}
\caption{(Color online) Computation of the map (\ref{MAP}) for $\tau_{0}=2$ and three different
values of $a$.} \label{fig6}
\end{figure}

To compute the map we have again combined numerical with analytical results by
numerically solving Eqs.~(\ref{MFmodel1B-dif})
and (\ref{IntMap}) using the analytic solution for $P_2^A(t;p_n)$.
Our results for $\tau_{0}=2$ are summarized in Fig.~\ref{fig6},
which clearly predicts that synchronization appears via a saddle node bifurcation.
For $a>a_{c}$ the function $f$ has no fixed points, so the system
is not able to synchronize.
At $a=a_{c}$ the function $f$ just touches the line
$p_{n+1}=p_{n}$, and for $a<a_{c}$ there appear two fixed points related to periodic
orbits. The lower fixed point corresponds to the stable oscillation that we observe
in our simulations; the other fixed point corresponds to an unstable limit cycle.
From this analysis we find that the critical point for $\tau_{0}=2$ occurs at $a_{c}=-1.42$.

\begin{figure}[h]
\begin{center}
\includegraphics[width =2.3in]{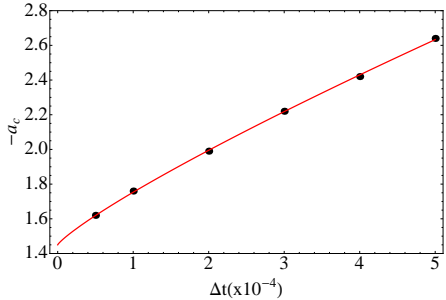}
\end{center}
\caption{(Color online) Variation of the critical point with the integration
step size $\Delta t$ (see text).  The continuous curve is
the best power law fit to the numerical data, $-a_c=0.303(\Delta t)^{0.846} +1.44$. Extrapolating
to $\Delta t \to 0$ leads to the critical point
$a_c=-1.44$, which is close to the value $a_c=-1.42$ obtained from the
saddle node analysis of the master equation.} 
\label{fig7}
\end{figure}

We can also estimate the critical point directly from the integral equation~(\ref{MFmodel1}). To
arrive at an ever more accurate critical point via this route, we must use ever smaller time steps
$\Delta t$ in the numerical integration.  In Fig.~\ref{fig7} we present the results as we decrease
the time step, and obtain the value $a_c=-1.44$ upon extrapolation to $\Delta t \to 0$.
The two methods to calculate the critical point thus show a high level of
agreement.

\section{Conclusions}
\label{sec5}

It is well known that a globally syncrhonized state can not occur in Markovian arrays of coupled
two-state systems.  A number of models of coupled arrays of two-state units have been constructed that
depart in one way or another from Markovian dynamics and that as a result do lead to global
synchronization. These models share a feature: the memory lies in the coupling between units and
does not affect the dynamics of the units themselves.

We have here constructed a novel departure from Markovian behavior that also leads
to globally synchronized dynamics, and that incorporates a physical feature endemic to two-state
systems: a refractory period. Specifically, in our model the single units are not Markovian
to begin with but instead, upon arrival at one of the two states, they are subject to
a delay time before they can return to the other state. The coupling among units is reflected in a
state dependence of this delay time.  In other words, the refractory period itself is affected
by the state of the system and therefore changes with time. 

In this globally coupled array we have found a synchronized state that exhibits highly anharmonic
dynamics with cyclical discontinuities.
The global dynamics of the network can be described by a generalized master equation.
We have solved the generalized master equation to show
that this type of synchronization is the result of a saddle node bifurcation in the
phase space of the system. The critical coupling strength, as well as the amplitude and
period of oscillations, are derived analytically by explicitly solving the generalized
master equation.
We find that as we approach the critical coupling parameter $a_c$ from below,
transient oscillatory behavior of longer and longer duration $T$ occurs, with $T\to\infty$ as
$a \to a_c$. We are currently exploring the functional dependence of $T$ on $|a-a_c|$.

The main points to be emphasized can be
summarized as follows.  Incorporating a refractory period that depends on the global state of the
system has given rise to a rich dynamics, one that can exhibit oscillations and
coexistence between the synchronous and asynchronous states (bistability).
The nature of the synchronization that leads to self-organization far from equilibrium is here
not related to the destabilization of the thermodynamic branch.  Indeed, this branch, which leads
to stationary behavior, remains stable and coexists with the stable oscillatory branch.  
This is an entirely different synchronization paradigm than that obtained from the usual Hopf bifurcation
and the attendant destabilization of the thermodynamic stationary state.
We have ascertained that our model is robust by considering two variations of the model and showing
that all three models lead to similar synchronization properties. 

Refractory properties are evident in a large variety of systems ranging from the
physiological (refractory properties of an excitable membrane) to the physical (blinking quantum
dots).  The rich dynamics of our simple on-off model makes this an excellent candidate
for adaptation to specific applications.

\section*{Acknowledgements}

DE thanks FONDECYT Project No. 11090280 for financial support.
UH acknowledges the start-up support (Grant
No.11-0201-0591-01-412/415/433) from Indian Institute of Science,
Bangalore, India.
KL gratefully acknowledges the NSF under Grant No. PHY-0855471.

\appendix
\section{Generalized master equation for mean field coupling}
\label{app}

In this Appendix we explore possible explicit functional forms of $\tau$ and the concomitant 
behavior of $\Theta$.  Note that we are ultimately especially interested in finding the
long-time behavior of the coupled array.
To make the case clear, we build up from the simplest possible scenarios to the most
complex that we will encounter.

\begin{figure}[]
\begin{center}
\includegraphics[width =2.7in]{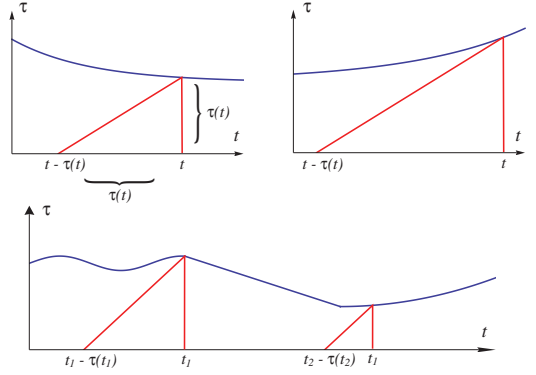}
\end{center}
\caption{
Left upper panel: Schematic of the construct leading to the generalized master equation
Eq.~(\ref{MFmodel3}) when $\tau(P_2(t))$ is analytic and decreasing with time. 
Right upper panel: The corresponding schematic when $\tau(t)$ is analytic and increasing
with time.  The master equation~(\ref{MFmodel3}) again results provided $d\tau/dt <1$.
Lower panel: again Eq.~(\ref{MFmodel3}) applies for continuous $\tau(t)$
even if there are discontinuities in its derivative, provided $d\tau/dt<1$.}
\label{schematic1}
\end{figure}

Consider first the case of a function $\tau(P_2(t))$ that decreases smoothly
with time, as sketched in the upper left panel of Fig.~\ref{schematic1}.
The construct in the sketch clearly shows that the \emph{inequality}
is valid only at the foot of the triangle.  Units that arrived earlier than
$t-\tau(t)$ will have left before time $t$ and thus do not obey the inequality.
The $\Theta$ function can then be expressed in the simple form
\begin{equation}
\Theta(t,t')=\theta \left(t' - t + \tau(P_2(t)) \right), 
\label{steadystate}
\end{equation} 
where $\theta$ is the Heaviside step function.  The resultant master equation reads  
\begin{equation}
P _{2}(t)= \int_{t-\tau(P_2(t))}^{t}dt' F\left(P _{2}(t') \right).
\label{MFmodel3}
\end{equation}
If $\tau(t)$ increases smoothly with time, then the same analysis is valid provided
$d\tau/dt \leq 1$. This is apparent in the schematic in the upper right panel of the figure. Indeed,
the master equation~(\ref{MFmodel3} is valid for non-monotonic continuous functions $\tau(t)$
even if there are discontinuities in its time derivative, provided $d\tau/dt \leq 1$, cf.
lower panel of Fig.~\ref{schematic1}.

In summary, as long as $\tau(t)$ is a continuous function of time and $d\tau/dt \leq 1$ we
find that Eq.~(\ref{steadystate}) holds, which in turn implies the master equation (\ref{MFmodel3}).
This is of course in general a complicated equation for $P_2(t)$, nonlinear and nonlocal in
time. 

\begin{figure}[]
\begin{center}
\includegraphics[width =2.7in]{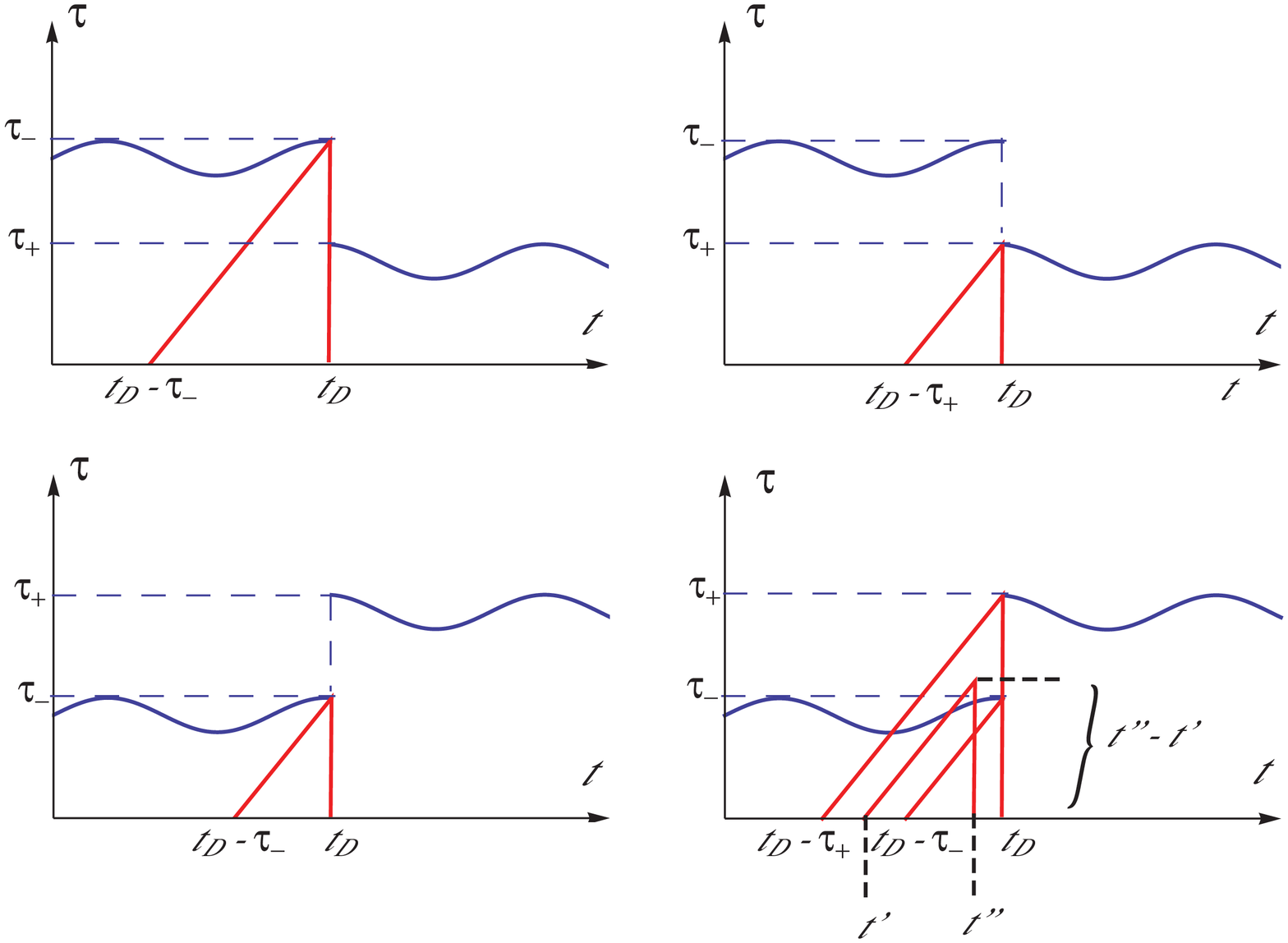}
\end{center}
\caption{Cases that can arise when $\tau(t)$ has discontinuities. Upper panels: high to low
discontinuities.  Lower panels: low to high discontinuities.
}
\label{schematic2}
\end{figure}

Next let us suppose that $\tau(t)$ itself has discontinuities. Two generic situations are possible,
as shown in Fig.~\ref{schematic2}. First consider a discontinuous abrupt decrease
of $\tau$ at $t\equiv t_D$ from a high value of $\tau_-$ to the left of the discontinuity to a low
value $\tau_+$ to the right, as seen in the upper panels. For later reference
we call this Case 1. 
The behavior away from $t_D$ on either side may be as described above, in which case it
requires no extra analysis, or increases with time and violates the condition $d\tau/dt <1$, a case
discussed subsequently.  In any case, we must be careful as we approach $t_D$.  The sketch
in the upper left panel of Fig.~\ref{schematic2} shows that approaching $t_D$ from below gives
\begin{equation}
\lim_{t\to t_D^-} \Theta(t,t') = \theta(t'-t_D+\tau_-).
\end{equation}
Approaching the discontinuity from above (right upper panel) leads to
\begin{equation}
\lim_{t\to t_D^+} \Theta(t,t') = \theta(t'-t_D+\tau_+).
\end{equation}
Thus we see that $\Theta$ inherits the
discontinuity of $\tau(t)$, but this is automatically ``built in" in
Eq.~(\ref{steadystate}), which is still valid. Hence the generalized master equation is again 
given by Eq.~(\ref{MFmodel3}).

Now suppose the discontinuity occurs in the opposite direction, from a low value $\tau_-$ to a high
value $\tau_+$ as we move from left to right.  This is shown in the lower panels
of Fig.~\ref{schematic2}. We call this Case 2. Again, sufficiently far from the discontinuity either
the previous discussion surrounding Fig.~\ref{schematic1} continues to hold, or the subsequent
discussion does.  As we
approach $t_D$ from the left all continues as before up to the value $t'=t_D - \tau_-$,
\begin{equation}
\lim_{t\to t_D^-} \Theta(t,t') = \theta(t'-t_D+\tau_-).
\end{equation}
This is clarified in the left lower
panel of Fig.~\ref{schematic2}.  However, when $t'$ lies in the interval $[t_D-\tau_+, t_D-\tau_-]$
then for $t"$ in the range $[\bar{t},t_D]$ lying between $t'$ and $t_D$
we see from the figure that the inequality is not satisfied,
i.e. that $t'-t' >\tau(t")$. Here $t'- \bar{t}=\tau(\bar{t})$.
However, when $t"$ continues to move up within the range $[t',t_D]$, the
inequality is once again satisfied, $t'-t < \tau(t")$.  The change from ``not satisfied" to
``satisfied" occurs at the value $t' = t_D-\tau_-$, so that
\begin{equation}
\lim_{t\to t_D^+} \Theta(t,t') = \theta(t'-t_D+\tau_-).
\label{before}
\end{equation}
Comparing the last two equations, we see that $\Theta$ is now in fact \emph{continuous} in spite of
the discontinuity in $\tau(t)$, that is, $\Theta$ now does not inherit the discontuity as it did
when $\tau(t)$ abruptly decreases. This asymmetry in behavior is of course a consequence of the
direction of the arrow of time. 

Furthermore, additional ``different" behavior is observed as we continue further upward in time.
Suppose we now look at values of $t$ in the next range, $[t_D,t^*]$, where $t^*$ satisfies
the equation 
\begin{equation}
t^* -(t_D - \tau_-) = \tau(t^*).
\end{equation}
The inequality is satisfied for all $t$ in this range provided that $t'-t < \tau_-$, that is,
\begin{equation}
\lim_{t\to t_D^+} \Theta(t,t') = \theta(t'-t_D+\tau_-).
\label{after}
\end{equation}
Comparing Eq.~(\ref{after}) with Eq.~(\ref{before}) reveals a remarkable behavior: provided
$t\in[t_D,t^*]$, $\Theta(t,t')$ does not depend on $t$!  Beyond $t^*$ the behavior as expected
from our analysis of a smooth function $\tau(t)$ is recovered.

\end{document}